A New, Long-Lived, Jupiter Mesoscale Wave Observed at Visible Wavelengths

Amy A. Simon[1,*], Ricardo Hueso[2], Peio Iñurrigarro[2], Agustín Sánchez-Lavega[2], Raúl Morales-Juberías[3], Richard Cosentino[1,4], Leigh N. Fletcher[5], Michael H. Wong[6], Andrew I. Hsu[6], Imke de Pater[6], Glenn S. Orton[7], François Colas[8], Marc Delcroix[9], Damian Peach[10], Josep-María Gómez-Forrellad[11]

1. NASA Goddard Space Flight Center, Solar System Exploration Division, 8800 Greenbelt Road, Greenbelt, MD 2077, USA
2. Física Aplicada I, Escuela de Ingeniería de Bilbao, UPV/EHU, Bilbao, Spain
3. New Mexico Institute of Technology and Mining, 801 Leroy Place, Socorro, NM 8780, USA
4. NASA Postdoctoral Program Fellow
5. Department of Physics & Astronomy, University of Leicester, University Road, Leicester, LE1 7RH, UK
6. University of California at Berkeley, Astronomy Department Berkeley, CA 947200-3411, USA
7. Jet Propulsion Laboratory, California Institute of Technology, 4800 Oak Grove Drive, Pasadena, CA 91109, USA
8. IMCCE, Observatoire de Paris, PSL Research University, CNRS-UMR 8028, Sorbonne Universités, UPMC, Univ. Lille 1, F-75014, Paris, France
9. Société Astronomique de France, Commission des observations planétaires, Tournefeuille, France
10. British Astronomical Association, Burlington House, London, UK
11. Fundació Observatori Esteve Duran, Barcelona, Spain

* Corresponding Author email: amy.simon@nasa.gov

Abstract

Small-scale waves were observed along the boundary between Jupiter's North Equatorial Belt and North Tropical Zone, ~16.5° N planetographic latitude in Hubble Space Telescope data in 2012 and throughout 2015 to 2018, observable at all wavelengths from the UV to the near IR. At peak visibility, the waves have sufficient contrast (~10%) to be observed from ground-based telescopes. They have a typical wavelength of about 1.2° (1400 km), variable-length wave trains, and westward phase speeds of a few m/s or less. New analysis of Voyager 2 data shows similar wave trains over at least 300 hours. Some waves appear curved when over cyclones and anticyclones, but most are straight, but tilted, shifting in latitude as they pass vortices. Based on their wavelengths, phase speeds, and faint appearance at high-altitude sensitive passbands, the observed NEB waves are consistent with inertia-gravity waves at the 500-mbar pressure level, though formation altitude is not well constrained. Preliminary General Circulation Model simulations generate inertia-gravity waves from vortices interacting with the environment and can reproduce the observed wavelengths and orientations. Several mechanisms can generate these waves, and all may contribute: geostrophic adjustment of cyclones; cyclone/anticyclone interactions; wind interactions with obstructions or heat pulses from convection; or changing vertical wind shear. However, observations also show that the presence of vortices and/or regions of convection are not sufficient by themselves for wave formation, implying that a change in vertical structure may affect their stability, or that changes in haze properties may affect their visibility.

1. Introduction

Small-scale waves, with 100 to 300-km wavelengths (~0.1 to 0.2° of longitude), have been observed on Jupiter multiple times, from the ubiquitous inertia-gravity waves seen by Voyager (Flasar and Gierasch 1986, Simon et al. 2015) to the equatorial gravity waves observed by Galileo (Arregi et al. 2009) or the Kelvin waves observed by New Horizons (Simon et al. 2015). However, another wave with larger wavelength (~1.2°, 1400 km), has also been observed in the North Equatorial Belt (NEB), near 16.5° N (planetographic latitude is used throughout, unless otherwise indicated). One such NEB wave was noted in a Voyager 2 image, but not noticed again until Hubble Space Telescope (HST) imaging in 2015 (Smith et al. 1979, Conrath et al. 1981, Simon et al. 2015). Simon et al. 2015 postulated that the NEB waves were formed as part of a baroclinic instability during cyclogenesis, however, on many previous dates cyclones were present, or had recently formed, without such waves evident.

Here we re-examine HST multi-color imaging data sets from 2012 to 2018, as well as the full Voyager 2 approach data. In addition, we report on NEB wave appearances in multiple visible and infrared ground-based data sets. In the past few years, the NEB waves have been observed sporadically, with life times of at least tens of hours, and are sometimes prominent enough to be seen in images acquired with ground-based telescopes. Section 2 summarizes these observational data. Section 3 describes the measurable aspects of the NEB waves. Preliminary General Circulation Modeling (GCM) is discussed in Section 4, with discussion of wave types and Earth analogies in Section 5.

2. Observations

Smith et al. (1979) showed a faint wave in Voyager 2 Imaging Subsystem Narrow-Angle Camera (NAC) data, reported in a single violet image on 3 July 1979 (see their Figure 3) in the NEB. Subsequent searches show the waves to be present almost every time this longitude region (~60° W System III) of the NEB was observed with sufficient spatial resolution, as shown in Figure 1. The waves are short, repeating, linear features visible against the background clouds. In the highest spatial resolution NAC images (~30 km/pixel), this NEB wave is much harder to spot, perhaps due to its low contrast. Near closest approach to Jupiter (9 July 1979), the Wide-Angle Camera (WAC) also offered a view of the NEB waves, with spatial resolution of ~115 km/pixel. Wave crests were identified in at least 107 Voyager 2 images from 26 June to 8 July 1979, and at all wavelengths observed, with highest contrast in the violet images (see Appendix Table A1).

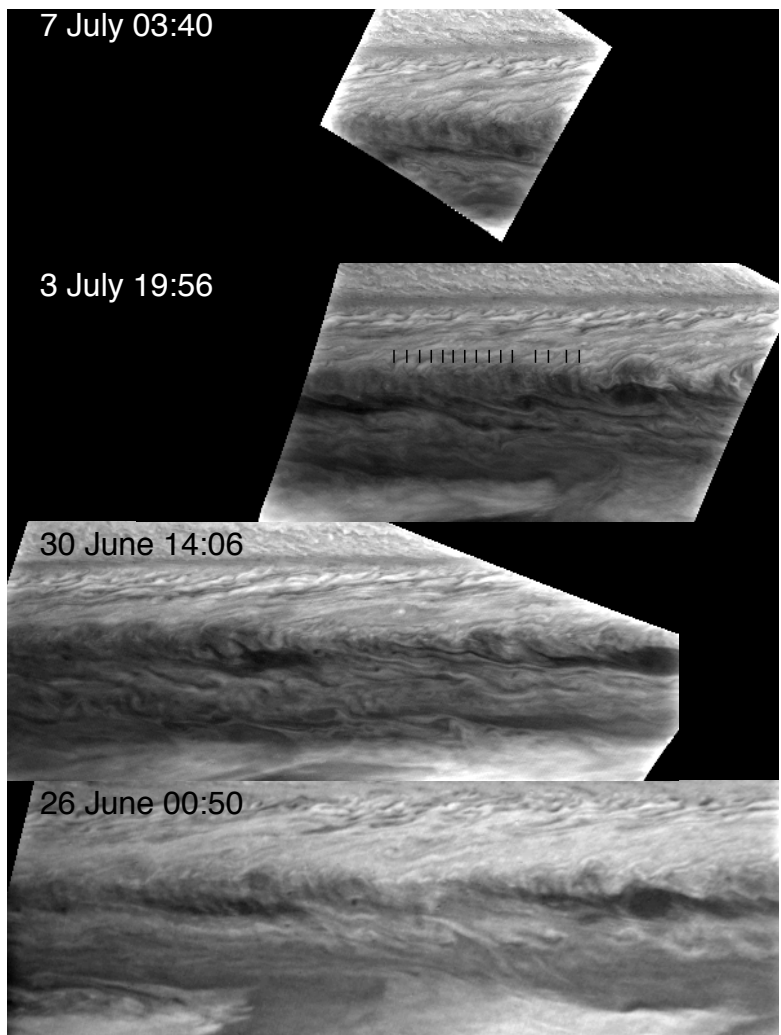

Figure 1. Voyager 2 time-sequence of NEB waves from the Imaging Subsystem NAC. Each map was produced at the same spatial resolution (0.1°/pixel) and centered near 16° N latitude, mapped over 20° in planetographic latitude and 60° in longitude (at closer range only part of that area is visible). A light unsharp mask was applied to increase feature visibility. The 26 June map was obtained in the green filter, the rest in the violet filter. The wave train is difficult to spot on 30 June, but evident on surrounding dates (lines added on 3 July to guide the eye).

A similar NEB wave was first noted, post-Voyager, in HST images in 2015, in part because of its extensive train of wave crests, and high contrast (Simon et al. 2015). In HST images 13 months later, no long wave trains were seen, and a small wave packet (~5° length) was only observed over one cyclone. However, in 2017, NEB waves could be spotted again over extensive sections of longitude. These waves were also observed at 5 μm for the first time, both from ground-based facilities and NASA's Juno spacecraft, as reported in companion papers (Fletcher et al. 2018, Adriani et al. 2018). A further search in older HST data sets reveal a faint NEB wave in 2012 images, as well (Appendix Figure A1). The appearance and longevity for both Voyager 2 and HST observations are noted in Table 1; dates without any observed NEB waves are shaded gray. Single hemisphere views (~80° to 260° W) in March 2017 also did not reveal any similar NEB waves.

Table 1. Summary of spacecraft observations

| Mission | Date | Sys III. Longitude (°) | Filters (nm)[2] | Related Features[3] | Visibility Notes |
|---|---|---|---|---|---|
| Voyager 1 | 1979-01-08 to 1979-03-05 | --- | --- | A, C, S | --- |
| Voyager 2 | 1979-06-26 to 1979-07-08 | 30-90 | 325, 400, 585, 615, 619 | FC. Some S to East, A & C elsewhere | ≥ 297 hours. |
| HST | 1995-02-17 1995-10-05 1996-10-21 2007-02-26 2007-03-26 2008-05-10 | --- | --- | A, C, S | --- |
| HST | 2012-09-20 | 295-300, 20-45, 130-160, 200-260 | 275 | A, C, FC?, S | Faint, ≥ 20 hrs. None at 763 nm. |
| HST | 2015-01-19 | 180-360 | 275, 343, 395, 502, 547, 631, 658, 889 | A, C, FC, S | ≥ 20 hrs |
| HST | 2016-02-09 | 50-55 | 343, 395, 467, 502, 547, 631, 658 | C. Elsewhere A, S. | ≥10 hrs. None at 275, 889. |
| HST | 2016-12-11 | 70-100 | 343, 395, 502, 631, 727, 750 | C. Elsewhere A, S. | Faint, ~10 hr. None in 225, 275, 889 |
| HST | 2017-01-11 | 85-135, 225-260 | 225, 275, 343, 395, 502, 631, 727, 750, 889 | C. Nearby | Faint at 225 and 889 |
| HST | 2017-02-01 | 25-75, 160-210 | 225, 275, 343, 395, 502, 631, 727, 750, 889 | C. Nearby A, S elsewhere. | ≥ 10 hrs. |
| HST | 2017-04-03 | 5-55, 235-305 | 275, 343, 395, 502, 547, 631, 658, 889 | A, C, FC. Some S to East | ≥10 to 20 hrs, but fading. |
| HST[1] | 2017-05-19 | 135-150, 190-215 | 343, 395, 502, 631, 727 | S, A and FC at ends. | Faint in 135-150 portion. None in 275, 889, faint in 727. |
| HST[1] | 2017-07-11 | 315-20 | 275, 343, 395, 502, 631, 727, 750 | A, C, FC. Weak S nearby? | Faint in 275. None at 889. |
| HST[1] | 2018-02-06 | 175-230 | 275 | S, FC?, nearby A | Faint in 275. None in 225, 343, 395, 502, 631, 727, 750, 889 |
| HST | 2018-04-01 | 180-230, 285-300 | 343, 395, 502 | C, A | Faint in 275, 631. None in 225, 889. |
| HST | 2018-04-17 | 35-180, 200-270 | 275, 343, 395, 502, 467, 631, 658, 889 | C, A, S | Faint at 225 and 889 |

Notes: 1. All observations included global coverage, except May and June 2017 and February 2018. 2. Filter wavelengths given here correspond to the names of WFC3/UVIS filters, as given in Table 6.2 of Dressel (2017). 3. Related features are those that are prominent over longitudes covered by the waves, or, on dates without waves, present at any longitude. A: Anticyclones, C: cyclones, FC: forming cyclones (less-defined shape), S: NEB storms.

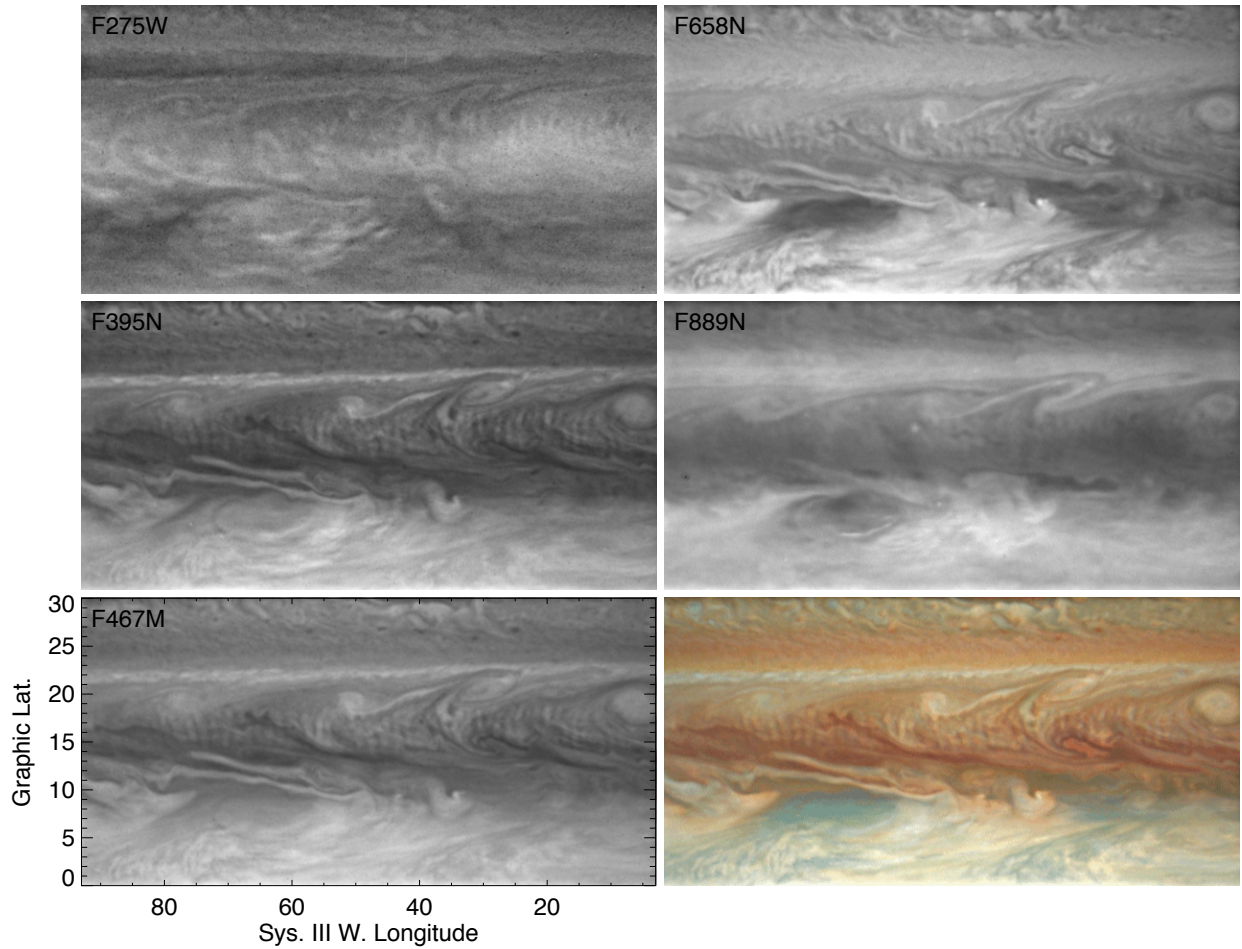

Figure 2. NEB wave crests in HST multi-wavelength images from 3 April 2017; the waves are just above center in each image. The color image is composed from images acquired with the filters F631N (red), F502N (green) and F395N (blue). Images have been enhanced with a light unsharp mask to bring out detail.

In 2017, NEB waves were also observable in ground-based observations, even using modest-size telescopes (0.36 to 0.5-m diameter). They were very prominent and well contrasted in observations obtained at the 1.05m telescope at the Pic du Midi Observatory (Figure 3). Table 2 lists the many times that wave features were observed. In some cases, a wave train was observed with enough continuity to be identified as the same feature, even if individual wave crests cannot be uniquely matched. Several wave trains were observed for more than 1 month. Appendix Figures A2-A4 show a selection of these ground-based images. NEB waves were observed again in 2018 with lower cloud contrast than in 2017 and also in locations close to the system of cyclones and anticyclones.

Table 2. Ground-based observations of NEB waves

| Observer | Date[1] | Sys III. Longitude (°) | Graphic Latitude (°) | Filter | Related features | λ (°) |
|---|---|---|---|---|---|---|
| P. Miles | 2017-02-22 | 209-234 | 15.9 ± 1.1 | IR700nm | --- | --- |
| D. Peach | 2017-02-26 (b) | 71-153 | 18.2 ± 1.1 | Visible | A | 1.7 ± 0.3 |
| P. Miles | 2017-03-05 | 100-116 | 17.8 ± 1.1 | IR700nm | --- | 1.5 ± 0.3 |
| D. Peach | 2017-03-07 (b) | 77-142 | 18.0 ± 1.1 | Visible | C, S | 1.9 ± 0.4 |
| A. Garbellini | 2017-03-12 (b) | 2-12 | 16.8 ± 1.2 | Visible | A, S | 1.6 ± 0.3 |
| C. Go | 2017-03-24 (b) | 7-57 | 17.0 ± 1.0 | Visible | A, C | 1.1 ± 0.2 |
| T. Olivetti | 2017-03-31 (c) | 23-57 | 17.2 ± 1.1 | Visible | A, S | 1.7 ± 0.4 |
| R. Bossman | 2017-04-08 | 60-67 | 15.7 ± 1.0 | Visible | A | 1.6 ± 0.4 |
| C. Go | 2017-04-10 (c) | 25-30 | 16.0 ± 0.5 | Visible | A, C, S | 1.5 ± 0.4 |
| M. Kardasis | 2017-04-15 (d) | 312-350 | 17.1 ± 0.9 | Visible | A, S | 1.4 ± 0.3 |
| C. Go | 2017-04-17 (d) | 318-351 | 16.7 ± 0.9 | Visible | A, C, S | 1.4 ± 0.5 |
| C. Go | 2017-04-19 (d) | 282-322 | 17.7 ± 0.4 | Visible | A, C, S | 1.4 ± 0.3 |
| T. Olivetti | 2017-04-21 (d) | 303-338 | 17.5 ± 0.6 | Visible | A, S | 1.6 ± 0.4 |
| C. Go | 2017-04-24 (d) | 302-325 | 17.5 ± 0.6 | Visible | C, S | 1.3 ± 0.4 |
| C. Go | 2017-04-26 (e) | 212-259 | 17.7 ± 0.7 | Visible | A, C, S | 2.2 ± 0.5 |
| A. Wesley | 2017-04-29 (d) | 281-294 | 17.9 ± 0.4 | Visible | A, S | 2.1 ± 0.4 |
| C. Go | 2017-05-01 | 244-275 | 17.3 ± 0.4 | Visible | A | 1.5 ± 0.5 |
| A. Wesley | 2017-05-03 (e) | 205-212 | 18.8 ± 1.1 | IR685nm | --- | --- |
| A. Wesley | 2017-05-05 (f) | 89-118 | 17.0 ± 1.7 | Visible | C, S | 2.0 ± 0.4 |
| C. Go | 2017-05-10 (e) | 199-213 | 17.0 ± 0.6 | Visible | A, S | 2.1 ± 0.4 |
| C. Go | 2017-05-18 (e) | 227-263 | 16.8 ± 0.4 | Visible | A, S | 1.9 ± 0.5 |
| Pic-Net | 2017-06-10 (f) | 62-285 | 17.3 ± 0.9 | Visible | A | 1.2 ± 0.1 |
| C. Zanelli | 2017-06-13 (f) | 90-113 | 17.0 ± 0.6 | Visible | A, S | --- |

Notes: 1. Letters (b) to (f) indicate systems that are identified in different dates. 2. Related features are as in Table 1. Appendix Figure A2 shows maps of the NEB on these images displaying the wave activity. The longitudinal range contains the extremes of the different wave systems that can be seen on different dates. In some cases, the images show fragmented waves that do not necessarily extend all the way between both limits.

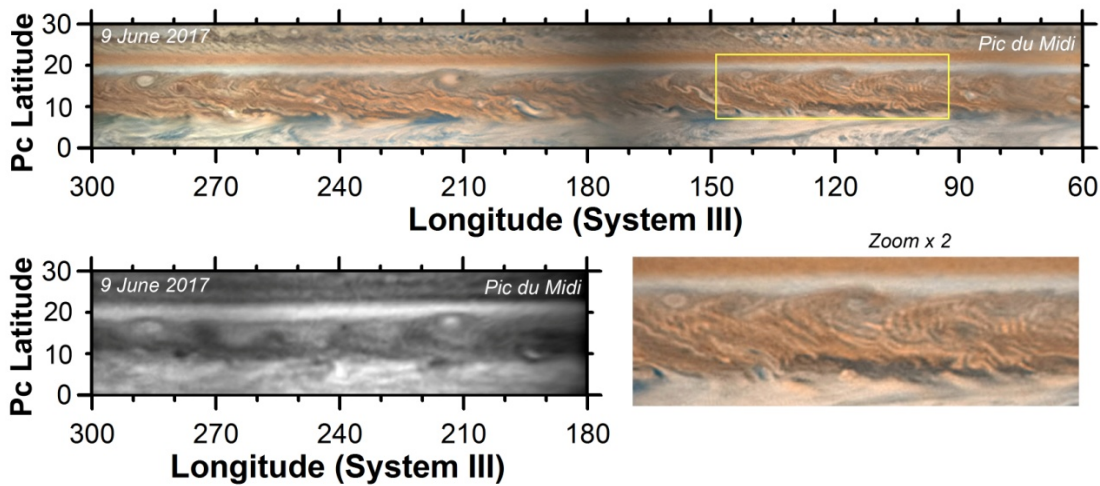

Figure 3. Ground-based observations of the NEB at visible wavelengths (top and bottom right) and in the methane absorption band (bottom left). In this image set, the NEB waves occur primarily in locations where high hazes are located; in the methane band an apparent chain of vortices forms a larger-scale wave with a dearth of haze opacity to the west of the smaller NEB waves (Fletcher et al. 2017, Fletcher et al. 2018) Note: 15° centric = ~17° graphic latitude.

3. Analysis

In Voyager 2 and HST images, the NEB waves have the most contrast at violet wavelengths, often with lower contrast at red wavelengths. When well-formed and observed at all the available wavelengths in HST data, the contrast is lowest at UV and methane absorption-band wavelengths, Figure 4. In Sept. 2012 and Feb. 2018, wave crests are seen only at UV wavelengths (275 nm), but not in visible to near-IR continuum or methane gas absorption bands. At a wavelength of 275 nm, Rayleigh scattering limits visibility (optical depth 1) to altitudes above 0.5 bar and absorption in the methane band at 890 nm produces optical depth 1 at ~ 0.1-0.2 bar. The waves have maximum contrast, ~10%, in violet-blue wavelengths and are separate from the background cloud tops (~1 bar) which have maximum contrast in the red and near-IR continuum. The NEB wave visibility is likely due to a combination of their altitude location at 0.1 – 0.5 bar, just above the cloud tops of the vortices, and aerosol/haze properties at short wavelengths. If the waves were located in the main clouds near 1 bar, the red and continuum bands would show maximum contrast, while waves in the stratospheric hazes would have maximum contrast in the UV.

The measurements of latitudinal extent depend on the wave contrast, but also that of the background cloud features; it is not always obvious where the wave crest begins or ends. Table 3 shows the average latitudinal extent of the waves, ~2.5°. The waves are largely aligned north-south, with slight westward tilts with latitude. The most extreme tilts are about 33° and may show curvature, usually near cyclones. For example, the wave over the cyclone in Feb. 2016, and over a cyclone near the east end of the wave train in Jan. 2015, Fig. A1. This can also be seen in the Pic-du Midi imaging data in Figure 3.

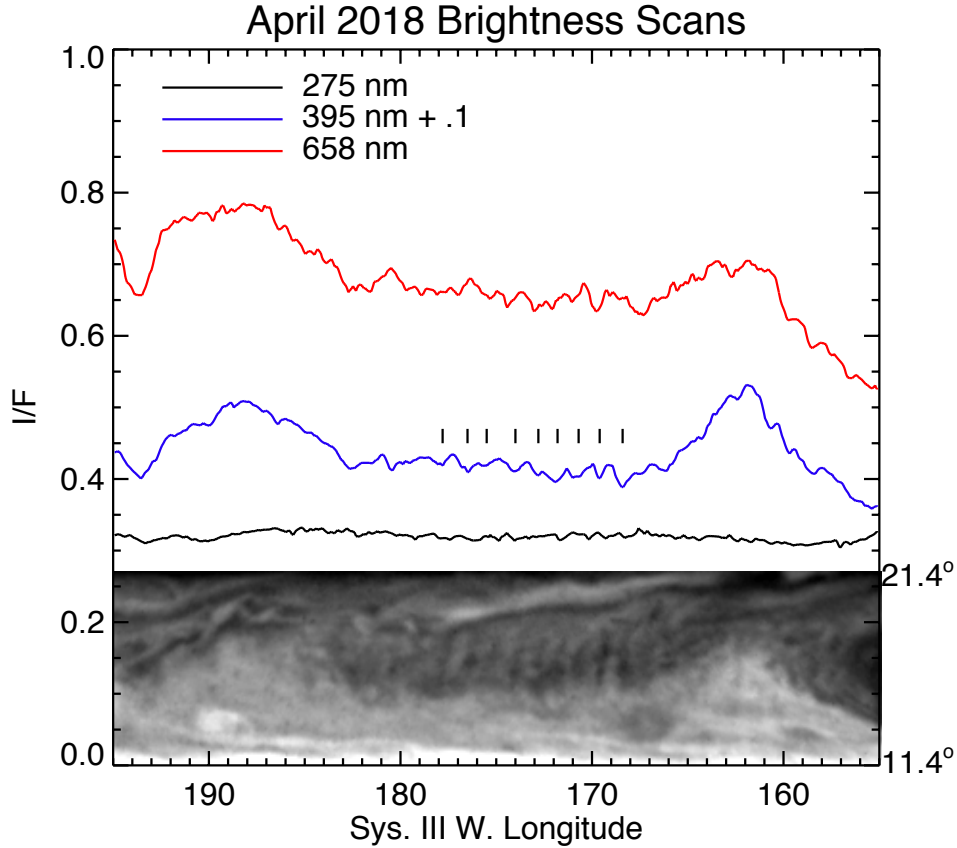

Figure 4. Brightness scans at 16.4° +/- 0.1° in HST images from 17 April 2018. White anticyclonic regions are seen near 162° W and 188° W in the image at 395 nm. Wave crests appear as periodic brightness variations above the background clouds; the 395-nm curve is shifted by 0.1 for clarity. Maximum contrast of the wave crests above the background features is ~3%, ~10% and ~6% at 275, 395 and 658 nm, respectively.

The central latitude of the NEB waves also varies slightly as seen in Tables 1 and 2, but much of this is likely due to the nearby vortices, which may dynamically shift the wave systems or at least mask their visibility. For example, in 2012 the average planetographic latitude is 17.7° ± 0.4° at longitudes away from any anticyclone, while for wave crests near the anticyclones it is 16.7° ± 0.3°, implying the anticyclones "shift" the NEB waves southward. In contrast, the cyclones push the wave crests, or their visibility, northward; wave crests over cyclones in July 1979 average 17.1° ± 0.3°, and in Feb. 2016 average 16.9° ± 0.6°, while in April 2017, they average 16.3° ± 0.4° away from any cyclones. Thus, the presence of cyclones or anticyclones accounts for much of the variation in central wave crest latitude in Table 1. Shifts of this nature are evident in Fig. 2, near the central anticyclone, and in Fig. A1 on the west end of the May 2017 wave train. The possible dynamic shift may be due to movement in the zonal flow (Johnson et al. 2018) or to a local change in the condensable haze/cloud that traces the waves near the vortices. Alternatively, it may be an observational effect caused by higher cloud tops in the anticyclones or cloud depletion above the cyclones at the vertical level of the waves, rendering them unobservable in the cloud fields.

The length of the wave trains is highly variable, from 5° in Feb 2016 (over a single cyclone), to covering most of a full hemisphere in 2015. However, the longitudinal wavelength has been nearly constant from the Voyager era to HST (Table 3) and in good agreement with ground-based data (Table 2). In the Voyager 2 data, a NEB wave is seen over at least 12 days, while the HST data sets are limited to approximately 20 hours of data, so a complete wave life cycle cannot be determined. Sometimes the NEB wave remains prominent over the full HST 20-hr observations, but on other dates it begins to fade over that time period. The amateur data sets provide evidence that some wave trains may last more than 30 days, though again, temporal sampling is incomplete.

Despite the many observations and multiple wave crests seen on all these dates, velocities are difficult to determine. In particular, it is impossible to conclusively identify the same wave crest from frame to frame with time separations of 10 hours, as aliasing may occur. In some cases, shorter time separations are available, but the motions are small and the uncertainties are large, leading to inconsistent measurements of wave crest behavior. The best data set for this is the Voyager 2 image set, because there were many images acquired over 6 days, some with ~1-hr separations, though several HST data sets have measurable motions, listed in Table 3.

Table 3. Measured NEB wave properties

| Date | Latitude (graphic) | Latitude Extent (°) | $\lambda$ (°) | Velocity, c (m/s) | Phase Speed, c-u (m/s) |
|---|---|---|---|---|---|
| 1979-06-26 to 1979-07-03 | 17.1 ± 0.3 | 2.2 ± 0.2 | 1.2 ± 0.2 | -24.7 ± 5 | -5.2 ± 8 |
| 2012-09-20 | 17.2 ± 0.8 | 2.5 ± 0.2 | 1.2 ± 0.3 | inconclusive | |
| 2015-01-19 | 16.1 ± 0.6 | 2.4 ± 0.6 | 1.1 ± 0.1 | inconclusive | |
| 2016-02-09 | 16.9 ± 0.4 | 2.4 ± 0.2 | 1.1 ± 0.1 | -17.5 ± 10 | -2.5 ± 10 |
| 2016-12-11 | 16.1 ± 0.1 | 2.3 ± 0.3 | 1.1 ± 0.1 | inconclusive | |
| 2017-01-11 | 16.5 ± 0.3 | 2.6 ± 0.5 | 1.2 ± 0.1 | -14.2 ± 10 | -4.2 ± 10 |
| 2017-02-01 | 16.5 ± 0.3 | 2.4 ± 0.3 | 1.1 ± 0.1 | Inconclusive | |
| 2017-04-03 | 16.3 ± 0.2 | 2.9 ± 0.5 | 1.2 ± 0.2 | -15 ± 15 | -3 ± 15 |
| 2017-05-19 | 16.5 ± 0.5 | 2.4 ± 0.3 | 1.1 ± 0.1 | --- | |
| 2017-07-11 | 16.2 ± 0.2 | 2.8 ± 0.4 | 1.1 ± 0.1 | --- | |
| 2018-02-06 | 16.2 ± 0.2 | 2.8 ± 0.3 | 1.3 ± 0.2 | --- | |
| 2018-04-01 | 16.5 ± 0.7 | 4.3 ± 1.2 | 1.2 ± 0.2 | --- | |
| 2018-04-16 | 16.0 ± 0.3 | 2.5 ± 0.5 | 1.1 ± 0.1 | Inconclusive | |

**Note:** Negative phase speeds indicate westward motion. For consistency, wave phase speeds are computed with respect to the zonal wind profiles previously measured for those dates (Simon-Miller and Gierasch 2010 for 1979, Tollefson et al. 2017 for 2016, Johnson et al. 2018 for 2017).

4. Modelling

On Earth, gravity waves are generated by flow over topography, by convective storms, by hurricanes, jets and fronts systems, and atmospheric dipoles (i.e., pairs of cyclones and anticyclones) (Nolan and Zhang 2017, Plougonven and Zhang 2014). On Jupiter, vortex interactions and adjustments, as well as convective heat pulses may also produce waves. In all observed cases in Table 1, cyclones and anticyclones are present, as well as convective regions. We used the Explicit Planetary Isentropic Coordinate (EPIC) GCM (Dowling et al. 1998) to explore some of these scenarios, focusing on the interactions between spots in these regions as a source of gravity waves.

We initialized a small domain model covering 60° in longitude (with 0.23 °/pixel resolution) and 25° in latitude, from 5° to 30° (with 0.25 °/pixel resolution). We used the zonal winds derived from Cassini observations (Porco et al. 2003), which for this region is indistinguishable within the measurement errors from HST and amateur observations throughout 2017 (Tollefson et al. 2017, Hueso et al. 2017, Johnson et al. 2018). The model has 18 vertical layers extending from 0.01 mbar to 7 bars and was initialized with a temperature profile and Brunt-Väisälä frequency, $N$, derived from Gemini/TEXES Observations acquired in 2017, Figure 5. For higher pressures, the temperature profile is extrapolated so that the value of the potential temperature decreases linearly with depth. The details of the calculation of the potential temperature in the model are explained in the appendix of Dowling et al. 1998, while the details of the calculation of $N$ can be found in the appendix of Dowling et al. 2006.

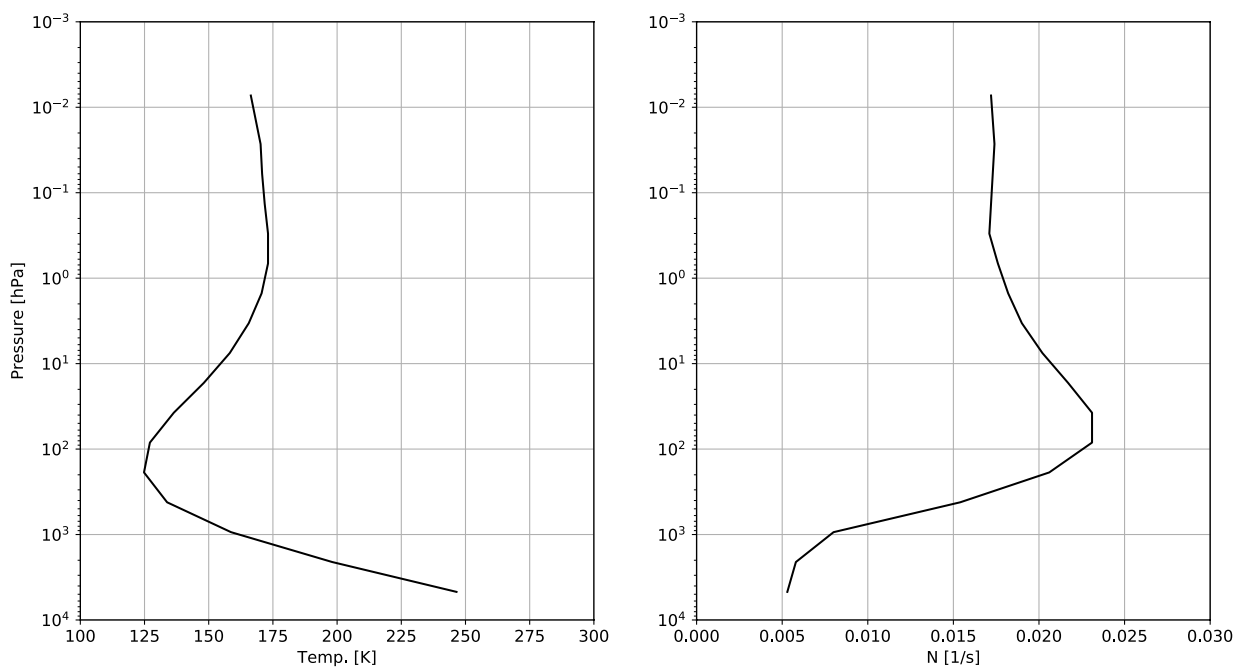

Figure 5. Derived Jupiter temperature profile (left) and Brunt-Väisälä, $N$, profile (right) used in the EPIC GCM simulations.

The model is initially in geostrophic balance and perturbations are added to simulate the presence of spots in this region. The EPIC model is initialized in geostrophic balance. Therefore, any perturbation added to an initially equilibrated state will spontaneously radiate gravity waves. These initial, transient, waves radiate concentrically away from the perturbation (i.e., they are neither parallel nor perpendicular to lines of constant latitude). Spot interactions such as mergers are also a source of gravity waves in the model and may better match the observed NEB waves. We explored three types of interactions, with properties of the resulting waves summarized in Table 4:

- Case 1: Merging anticyclones, with cyclones also present in the domain. The spots are initialized with a size of 3° diameter and an initial rotational amplitude of 40 m/s. After approximately 14 simulated days, two of the anticyclones merge, producing wave trains similar to the wave trains in the observations (Fig. 6 top panel and Fig. 7 left panel).
- Case 2: Merging anticyclones, with no cyclones present. The spots are initialized with the same properties as in Case 1. After approximately 35 simulated days, the anticyclones merge producing wave trains (Fig. 6 middle panel and Fig. 7 middle panel).
- Case 3: Merging cyclones. The spots size and amplitude are initialized as in the previous cases. After approximately 20 simulated days, the two cyclones merge producing wave trains similar to those observed (Fig. 6 bottom panel and Fig. 7 right panel).

The waves produced in the simulations show morphological similarities to the observed NEB waves. Those forming off the cyclones show curvature, similar to those in Figure 8, top and bottom. In the model, the passage of an anticyclone to the north of the cyclones appear to make the wave trains more stable, reducing the tilt of the crests, and shifting the wave trains north and south, similar to Fig. 8, all three panels.

The simulations produce potential vorticity patterns that resemble the observed waves in terms of wavelength, speed, and lifetime, Table 4, but the HST imaging data directly observe aerosol opacity rather than potential vorticity. Models show that cloud opacity maps can differ significantly from potential vorticity maps (Marcus 2004, Palotai et al. 2014). Although our simulations imply that the waves may be commonly produced by vortex interactions, details of cloud microphysics may play a role in determining whether the waves are visible in imaging data.

Table 4: Summary of Observation Constraints and Model Output

| Parameter | Avg. Spacecraft Observations | Case 1 | Case 2 | Case 3 |
|---|---|---|---|---|
| Central Graphic Lat. | $16.5 \pm 0.5$ | 17 | 17 | 17 |
| Lat. Extent (°) | 2.5 | 2 | 1.4 | 2 |
| Wavelength (°)[3] | 1.2 | ~1 | ~1 | ~1 |
| Long. Extent (°)[1] | ~5 to >50 | 8 | 4 | 4 |
| Speed (m/s) | $-15 \pm 15$ | -17.5 | -7 | |
| Lifetime (Days)[2] | ≥12 | 20 | 8 | 20 |

Notes: 1. The length of individual wave trains is difficult to distinguish because of the background cloud features. 2. The lifetime observed in amateur data may exceed 30 days. 3. In the simulations the wavelength changes as a function of time as seen in Fig. 7

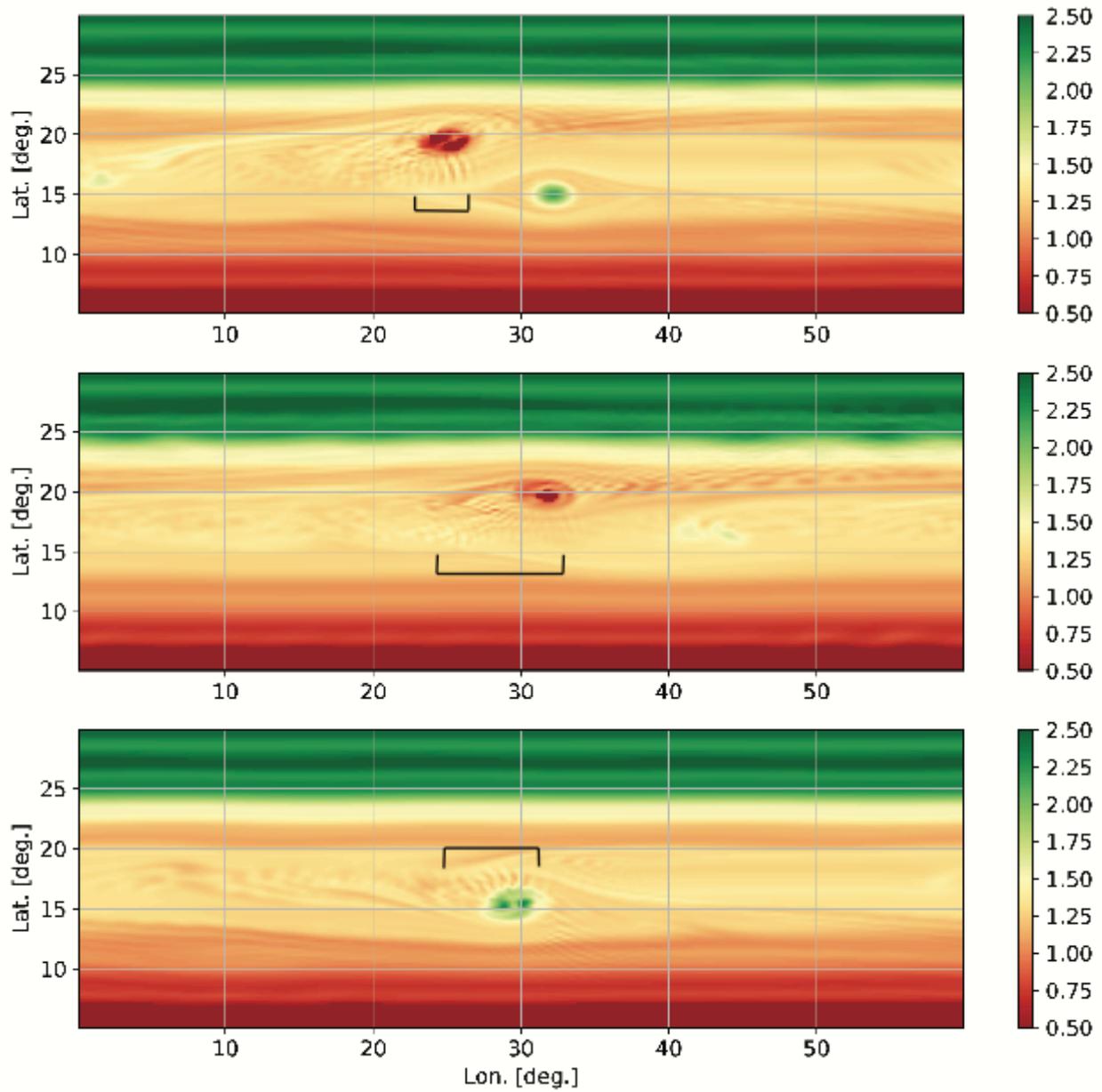

Figure 6. Potential vorticity maps for cases #1 (top), #2 (middle) and #3 (bottom) at days 17, 39, and 23 simulated days respectively. Green indicates cyclonic vorticity, while red is anticyclonic. The produced wave trains are marked.

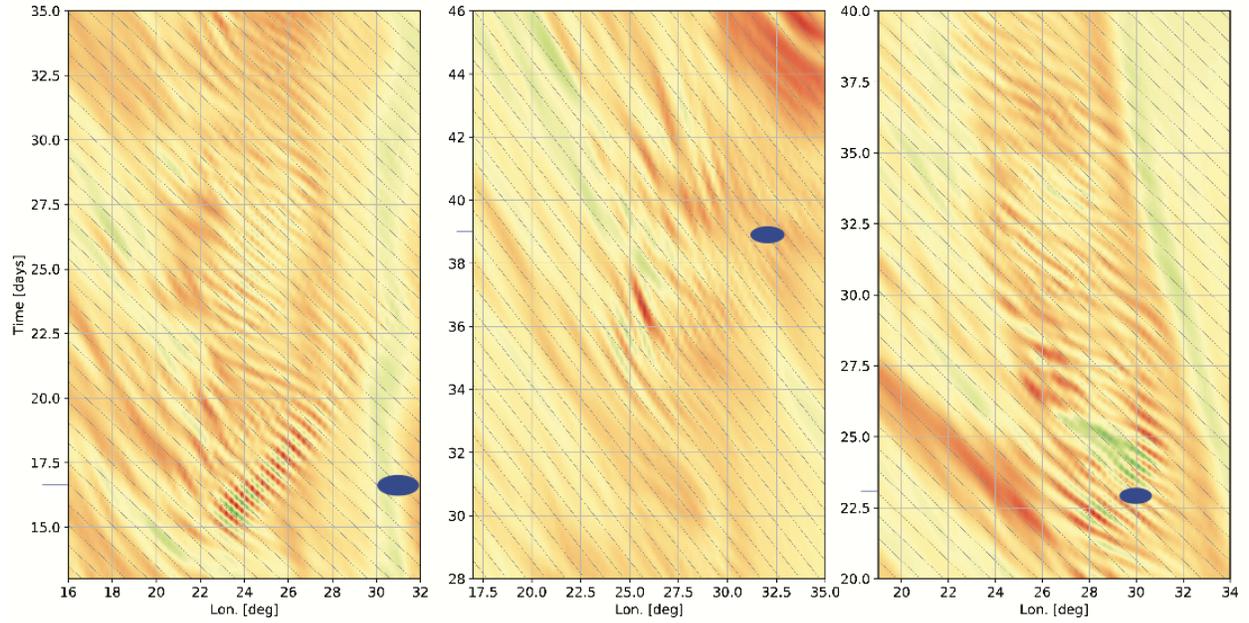

Figure 7. Temporal evolution of the potential vorticity at 17° latitude for cases #1 (left), #2 (middle) and #3 (right). These Hovmoller diagrams highlight the difference from the mean flow, and the location of the vortices and day of simulation in Fig. 6 are marked in blue. The gray diagonal lines are spaced by 1°, and their slope corresponds to a velocity of ~17.5 m/s.

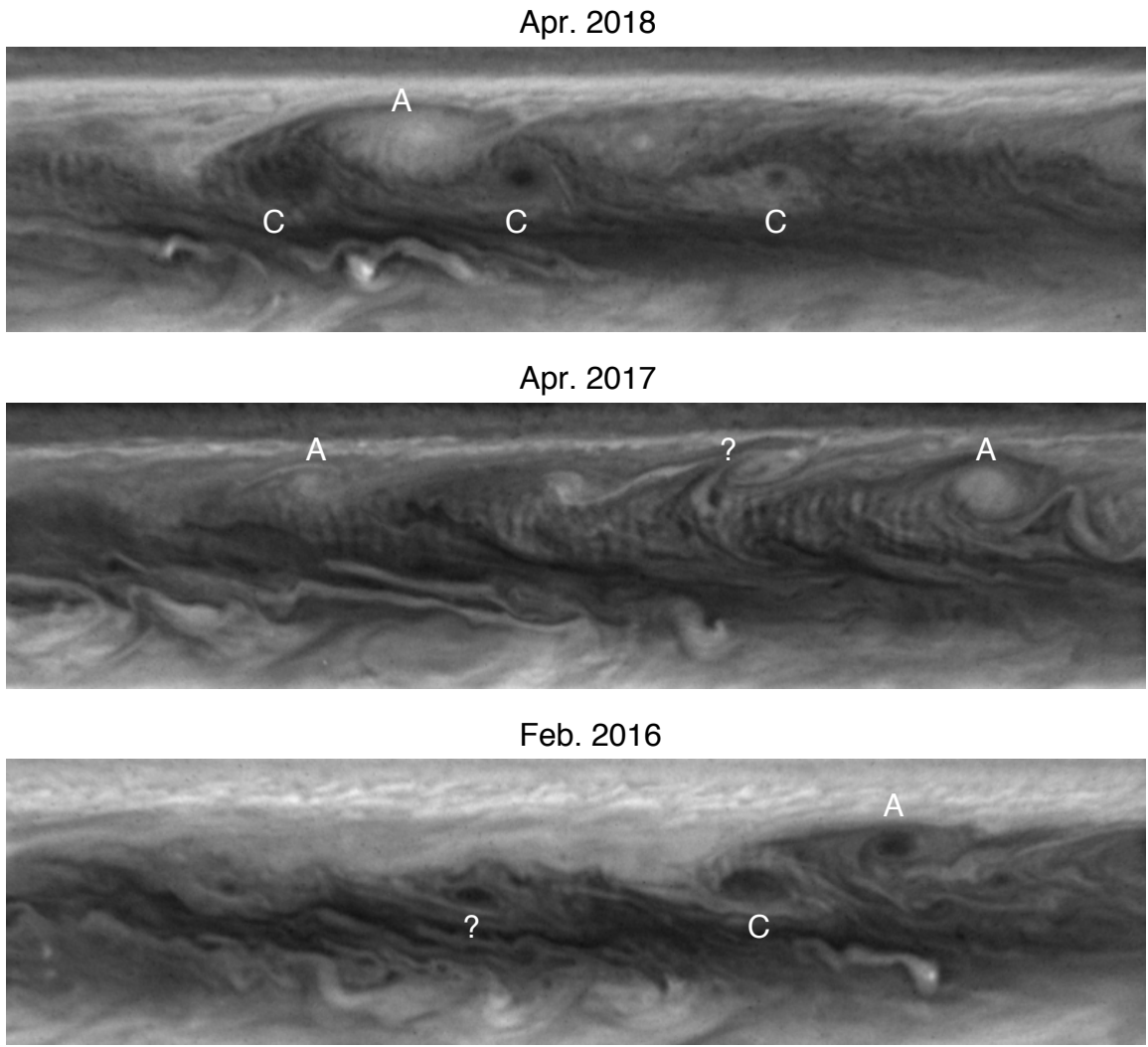

Figure 8. HST maps from April 2018, April 2017 and Feb. 2016 spanning 80° of longitude and 20° of latitude. These three dates roughly match the three cases shown in Figs. 6 and 7. Note that extensive wave trains are seen in Apr. 2018 among cyclones (C) and anticyclones (A). The Apr. 2017 map segment mainly shows anticyclones and features suggestive of a vortex (denoted by ?). In Feb. 2016, only a small wave train is seen over a single cyclone.

5. Discussion

5.1 Wave Formation Mechanisms
Most of the observed NEB wave trains begin on the westward side of cyclones or anticyclones, though not exclusively; all observed NEB wave trains has some sort of vortex present, although sometimes the wave trains extended far from the vortices. Based on the spacecraft observations, they are tightly confined to latitudes of 16.5° +/- 0.5° and have regularly spaced crests ranging in number from 6-12 or more, depending on where one defines the boundaries of an individual wave train. Numerical studies of Earth's atmosphere have shown that inertia-gravity waves (IGWs) should be generated in GCMs from the interaction of cyclones and anticyclones (Wang et al. 2009); IGWs are buoyancy waves and include the effects of the Coriolis forces, while pure gravity waves (GW) do not.  Small-scale IGWs are common in fluid flows and on Earth, caused by velocity shears, topography, thunderstorms, and geostrophic adjustment (e.g., Fritts 2015). Williams et al. (2003) used modeling and laboratory studies to show that spontaneous adjustment radiation can also form small-scale waves that interact with the baroclinic flow to selectively produce longer wave modes.

Cyclones on Earth are also known to form several other types of atmospheric GWs that could be analogous to the ones seen in this study. First, spiral density or buoyancy waves are observed to be radiating from the center of powerful hurricanes (Nolan and Zhang 2017). These very small horizontal waves have only been observed in the clouds of the hurricanes themselves, not outside of the storm like the Jupiter waves at 17° N. Earth cyclones can also produce small-scale GWs that radiate away from the storms as they experience baroclinic instabilities (Vallis 2006, Yue et al. 2014). The terrestrial waves also have very small horizontal wavelengths and are only observed in close proximity to the storm in fully, or partially, concentric rings that spread with distance.  Lastly, large-scale GWs can be produced from terrestrial cyclones as they go through periods of geostrophic adjustment caused by the 'obstruction effect' of local topography. On Earth, these large-scale GWs have large horizontal wavelengths (~500 km), are observed at greater distances away from the storms (see Figure 3 of Yue et al. 2014), and propagate upwards into the stratosphere. These wave trains are not the same scale or length as Jupiter's waves, in part because the tropical cyclones are subject to topography, latitudinal drift, and atmospheric conditions that are quite different from conditions in Jupiter's NEB.  Nonetheless, terrestrial tropical cyclones show a potential wave formation mechanism and if conditions allow (such as wave trapping), the wave train can grow.

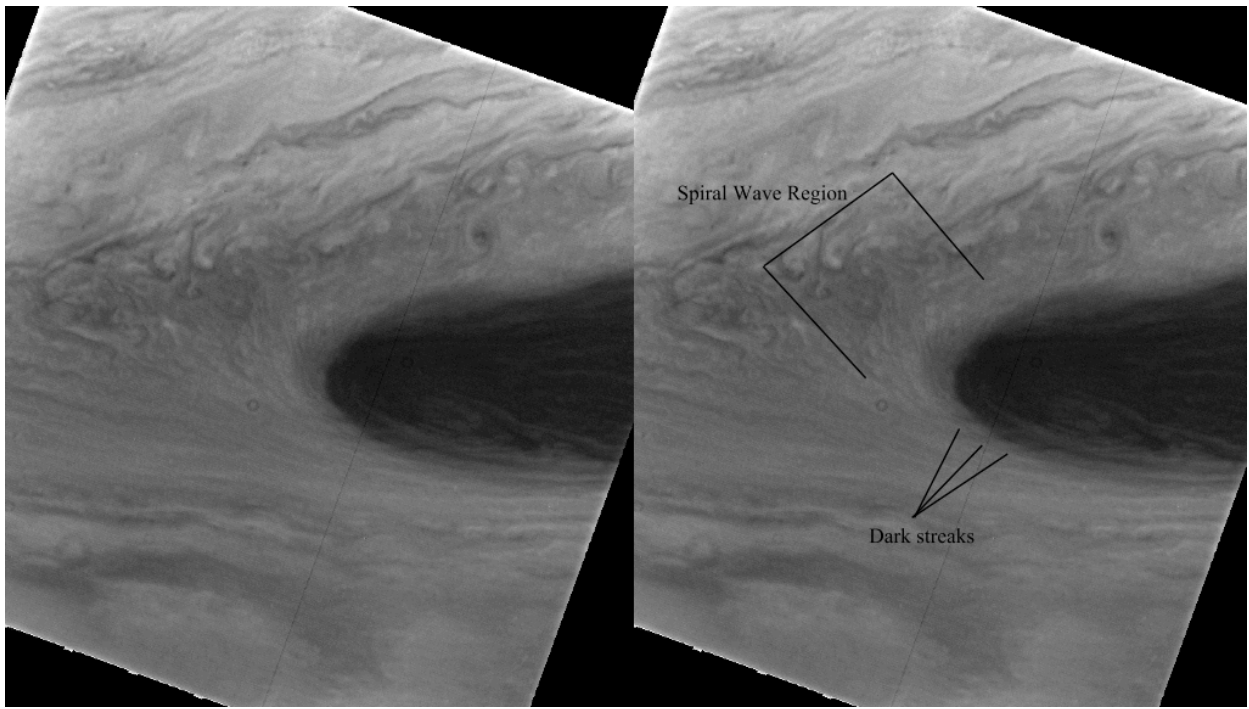

Figure 9. Voyager 1 closest approach image of a brown barge cyclone. Acquired on 4 March 1979 at 12:36:36, there is evidence of wave-like structures on the western periphery of the vortex. Potential wave features are annotated in the right panel. Image resolution is ~ 14 km/pixel.

For Jupiter, the smallest-scale spiral and baroclinic GWs would be below the ~175-km spatial resolution limit of HST. Very fine scale waves are observed in the NEB (and other regions) in recent JunoCam data with wavelengths of 55 to 174 km (e.g., Sanchez-Lavega et al. submitted). The highest spatial resolution Voyager 2 data did not include any images of the cyclones near the NEB waves. However, Voyager 1 did observe a large well-formed cyclone at a resolution of ~14 km/pixel (Figure 9) with a fine-scale structure suggestive of a spiral wave or GW, as well as a few near-horizontal dark streaks with spacings of 250 to 425 km and lengths of 500 to 750 km; similar streaks are seen on many Voyager vortices. These features could be part of the formation cycle of the NEB waves seen during Voyager 2 a few months later, but we cannot rule out that they may be unrelated.

However, none of these proposed formation mechanisms currently explain why NEB waves are only observed near some longitudes and are temporally variable. Without constant high-resolution monitoring, it would be next to impossible to connect observed individual wave crests to a specific cyclone or anticyclone, cyclone/anticyclone interaction, or some other sporadic process, even with the combination of spacecraft and ground-based coverage in 2016 to 2018. It is possible that ALL of these processes are producing individual NEB waves on different dates and longitudes, with the wavelength and location governed by atmospheric properties. If such properties change over time, waves may not be able to form, or may not be observable if present.

5.2 Wave Property Analysis
Initial studies had concluded that the NEB waves might be a baroclinic instability (Conrath et al. 1981, Simon et al. 2015), However, this interpretation is problematic for several reasons. First, such an instability would grow with longitude, and there is no evidence of changing wavelength over the longer wave trains. Second, Conrath et al. (1981) showed that the observed wavelength corresponds to a Rossby deformation radius of ~400 km, and this particular wave mode should be confined to the upper troposphere. However, that deformation radius leads to a Brunt-Väisälä frequency, $N$, of ~$1.8 \times 10^{-3}$ s$^{-1}$, which only occurs below the cloud deck (Fig. 5, right). In the fastest growing mode, the vertical scale of the waves would be very small, a few meters at most, too small to produce the observed contrast. Thus, it is hard to reconcile the waves with their observed properties above the cloud deck. Similar analysis of a Rossby wave, which could match the small phase speeds, also produces waves with very small vertical extent, a few meters or less.

As the simulations in Section 4 show that it is easy to generate IGWs/GWs, we now compare the observed NEB wave properties with the properties of an IGW or GW. For an IGW or GW in the Boussinesq approximation for a rotating planet, density variations are neglected except in the buoyancy term, and phase speed has components in both the horizontal or vertical direction, though the vertical component is often small (Dunkerton 2015).

The full IGW dispersion relation, which includes Coriolis forces, is given by (Fritts 2015):

$$\omega - uk = \sqrt{\frac{N^2 k^2 + f^2 m^2 + f^2/(4H^2)}{k^2 + m^2 + 1/(4H^2)}}$$

where $k$ is the horizontal wavenumber ($2\pi/1400$ km), $m$ is the vertical wavenumber, $N$ is the Brunt-Vaisala frequency, $H$ is the atmospheric scale height, $f$ is the Coriolis parameter, and $\omega$ is the wave frequency. For pure GWs, the Coriolis terms disappear, and if vertical wavelength $\ll 4\pi H$ (~315 km for H = 25 km), the $1/(4H^2)$ terms can be neglected (Holton and Alexander 2012, Lane 2015, Dunkerton 2015).

At 16.5° latitude, $f$ is ~$1 \times 10^{-4}$ s$^{-1}$. As the NEB waves are observed near 500 mbar, we use a value of $N = 1.1 \times 10^{-2}$ s$^{-1}$ (Fig. 5) and the observed horizontal velocity and wavelength to simultaneously solve for vertical wavenumber, $m$, and the vertical trace velocities, $w_T$:

$$u_T = u + \frac{1}{k}\sqrt{\frac{N^2 k^2 + f^2 m^2}{k^2 + m^2}}$$

$$w_T = \frac{1}{m}\sqrt{\frac{N^2 k^2 + f^2 m^2}{k^2 + m^2}}$$

Assuming the $u_T = c = -3.7$ m/s (Table 3 average), the IGW has $m^2 < 0$ and does not propagate vertically, while a pure GW has a vertical wavelength of ~300 m. If the full phase speed uncertainty is considered, vertically propagating 400-m or 700-m IGWs are also possible solutions. These yield vertical velocities of a few m/s, for the propagation GWs and IGWs. This would imply that the waves should be quite visible at the higher altitudes, which is not observed in the UV or methane-band filtered images (225, 275 and 889 nm). Thus, the non-propagating IGWs would be the preferred solution.

Alternately, the NEB waves may form deeper and vertically propagate above the cloud deck, possibly breaking at a critical level, or becoming ducted above the clouds. If they form below the clouds, near the water layer at 2 to 4 bars, they must vertically propagate a few scale heights to be visible near 500 mbar. At this depth the phase speed is not known, but if we assume the same full range of phase speeds above and $N = 1.8 \times 10^{-3}$ s$^{-1}$ (Conrath et al. 1981, Simon et al. 2015), this yields solutions for a GW with 1.9 to 4.5-km vertical wavelength and ~5 to 30-cm/s vertical velocity, or an IGW with wavelength of 2.5 km and 17 cm/s vertical velocity. These waves would reach the 500-mbar altitude in about 100 hours. Thus, either GWs or IGWs at this depth are also a plausible solution. It is not clear from the observational constraints if the NEB waves form near 500 mbar, or if they form much deeper and propagate vertically to this pressure. Indeed, 5-micron imaging of these waves, and similar analyses from Cassini CIRS temperature inversions also cannot uniquely constrain the wave formation altitude (Fletcher et al. 2016, 2018).

Lastly, for waves to form, the atmosphere must have the proper static stability. One measure of this is the Richardson number, $Ri$, defined by:

$$Ri = \frac{N^2}{(\partial u/\partial z)^2}$$

where $\partial u/\partial z$ is the vertical wind shear. For $N^2 = 0$ (and $Ri = 0$), the atmosphere has neutral stability and no oscillations result. If $N^2 > 0$, stable oscillations can result with higher values inhibiting vertical displacement, and if $N^2 < 0$ this increases the vertical displacements. In the resulting $Ri$, at $Ri < 0.25$ wind shear dominates and Kelvin-Helmholtz instability (turbulence) may arise, while for $Ri > 0.25$ stability generally dominates, with $N^2$ determining the oscillation amplitude, damping at higher values ($N^2 > 1$) (Young 2015).

The vertical wind structure in this region of Jupiter's atmosphere cannot be directly observed, but it can be inferred from the thermal wind equation (Gierasch et al. 1986, Simon-Miller et al. 2006, Li et al. 2006). However, retrievals of the vertical temperature profile from IR data are relatively insensitive to the temperatures below the nominal cloud deck. Above the cloud deck $Ri$ grows to $\gg 1$, which would indicate waves are damped and should not be forming, though it is possible that the active NEB convection is affecting the local stability.

5.3 Wave Temporal Appearance
Any waves that do form may be damped over time or break in the presence of vertical wind shear and critical layer absorption, depositing their energy and disappearing (Lane 2015). If conditions are no longer conducive, no new waves appear. The temporal variability of the NEB waves implies either the static stability is temporally variable, or else ever-present waves are simply not visible due to cloud/haze changes. While we do not know the exact time between the end of the Voyager 1 data and when the 1400-km NEB waves first appear, their presence in the earlier Voyager 2 data indicates that onset of wave formation is < 100 days.

It is of note that the 2015 NEB waves appeared in the same seasonal cycle as those seen during Voyager (northern autumnal equinox) and that identical waves have not been observed at similar southern latitudes. First, the structure of the zonal winds at equivalent North and South latitudes is very different and include the presence of the Great Red Spot. Additionally, the combination of Jupiter's orbital eccentricity (e= 0.0489) and rotational axis obliquity (3.13°) gives rise to more heating in the north and there has been a marginal detection of a seasonal component to haze thickness or brightness in the UV (Simon-Miller and Gierasch 2010); thicker haze may allow the waves to become more prominent. Unfortunately, there were no high spatial resolution imaging data near intervening northern equinoxes (1991, 2003) to confirm this hypothesis.

Alternately, numerical modeling of the wave-driven Quasi-Quadrennial Oscillation (QQO) showed that forcing a strong equatorial jet to descend in time produces smaller, weaker jets near +/-15° latitude in response (Cosentino et al. 2017). These jets vary in altitude and amplitude and could change the vertical wind shear in the region above the NEB cyclones. An alignment of the QQO cycle to the seasons may have a greater cumulative effect on the vertical temperature structure and/or haze thickness in the NEB. However, in the current epoch, the NEB waves have persisted longer than the QQO cycle, and we do not know how long they persisted after Voyager 2. Further observations of thermal and vertical wind structure, as well as near-UV brightness, are clearly needed for periods with and without NEB waves to further elucidate any differences that may be present. Future detailed GCM simulations will include a larger domain space with more realistic boundaries, damping, and exploration of parameter space (temperature profile, wind profile, Brunt-Vaisala frequency) to understand under which conditions they form and best match the observations, including wave train length, motions, and wave crest tilts. We will also explore the effect of different types of perturbations such as heat pulses and planetary-scale waves in producing NEB waves.

6. Conclusions

The large visible wavelength data set presented here gives several new insights on a new class of mesoscale waves observed at ~16.5° +/- 0.5° latitude. Their wavelength is a nearly constant 1.2°, and they are observed near both cyclones and anticyclones, sometimes shifting in latitude as they pass a vortex. The observed lifetimes of individual wave trains vary, but can exceed 30 days, and, when at maximum contrast, they are observable in even modest ground-based telescopes. The observed NEB wave properties are more consistent with inertia-gravity waves than Rossby waves or baroclinic instability, though the latter two cannot be entirely dismissed. Although the NEB waves appear at or above the cloud deck, the altitude of formation is not well constrained.

The presence of the NEB waves in current observations of Jupiter remains perplexing for several reasons. This region is consistently observed to have frequent convective outbreaks, accompanied by several interacting and sometimes merging cyclones, all of which are sources of atmospheric waves. As such, there is no shortage of processes to generate the NEB waves, but why are they so prevalent now and not over the previous 20 years? What has changed in this region for the waves to be present in Voyager-era images, disappear and then now re-appear? Is their presence predicated on the numbers of cyclones, an indication of some other slowly varying atmospheric condition like vertical wind shear, or on variations in tropospheric haze opacity? Modeling can reproduce IGW/GWs from vortex interactions, but further exploration of parameter space, including the effects of heat pulses to mimic convective outbreaks, is need to determine why they do not form at other times or latitudes. More intensive numerical modeling could also help determine which of these is more likely to reproduce the observed NEB wave train lengths. Further visible and infrared observations of the NEB, with and without waves, are also needed to understand subtle changes in the local wind, temperature, and cloud/haze structure.

Acknowledgements:
This work used data from the NASA/ESA HST Space Telescope, and A.A.S, M.H.W, G.S.O were supported by grants from the Space Telescope Science Institute, which is operated by the Association of Universities for Research in Astronomy, Inc., under NASA contract NAS 5-26555. These observations are associated with programs GO13067, GO13937/14334/14756/15262, GO14661, and GO14839. Jupiter maps are available at https://archive.stsci.edu/prepds/opal/ and https://archive.stsci.edu/prepds/wfcj/. R.G.C's research was supported by an appointment to the NASA Postdoctoral Program at the NASA Goddard Space Flight Center, administered by Universities Space Research Association under contract with NASA. R.H., P.I. and A.S.-L. were supported by the Spanish MINECO project AYA2015-65041-P with FEDER, UE support and Grupos Gobierno Vasco IT-765-13. P.I. also acknowledges a PhD scholarship from Gobierno Vasco. L.N.F. was supported by a Royal

Society Research Fellowship and European Research Council Consolidator Grant (under the European Union's Horizon 2020 research and innovation programme, grant agreement No. 723890) at the University of Leicester. Observations at the Pic du Midi observatory were acquired by the Pic-Net team, F. Colas, M. Delcroix, E. Kraaikamp, R. Hueso, D. Peach, C. Sprianu, G. Therin, with funding from Europlanet 2020 RI, which has received funding from the European Union's Horizon 2020 research and innovation programme under grant agreement No. 654208. We thank Dr. A. Ingersoll for a thorough review and useful comments.

Appendix A: Large Data Set Information

Several larger data sets were analyzed in this work. Figures A1-A4 provide full resolution views from several of the Hubble data sets. Ground-based images are shown in Figures A2-A4. Table A1 includes a list of Voyager images were NEB waves were identified.

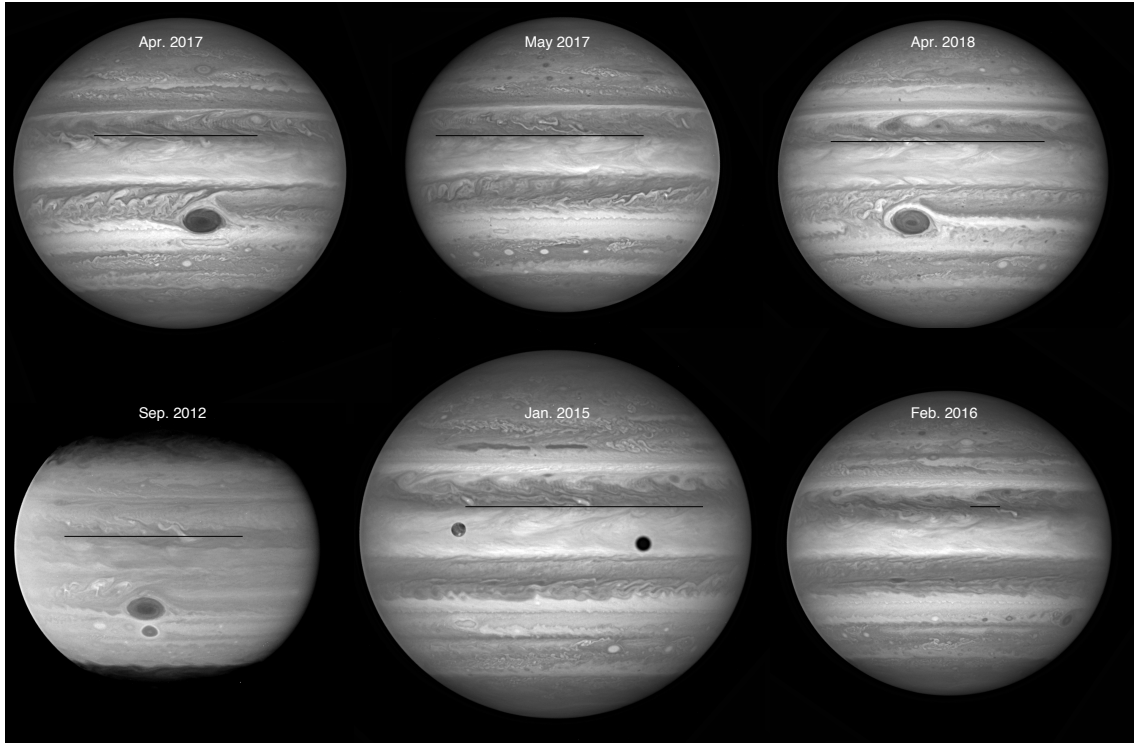

Figure A1: Full disk views of Jupiter from Hubble. Waves are delineated by the solid line. All images were acquired at 395 nm, except Sept 2012 which, is 275 nm.

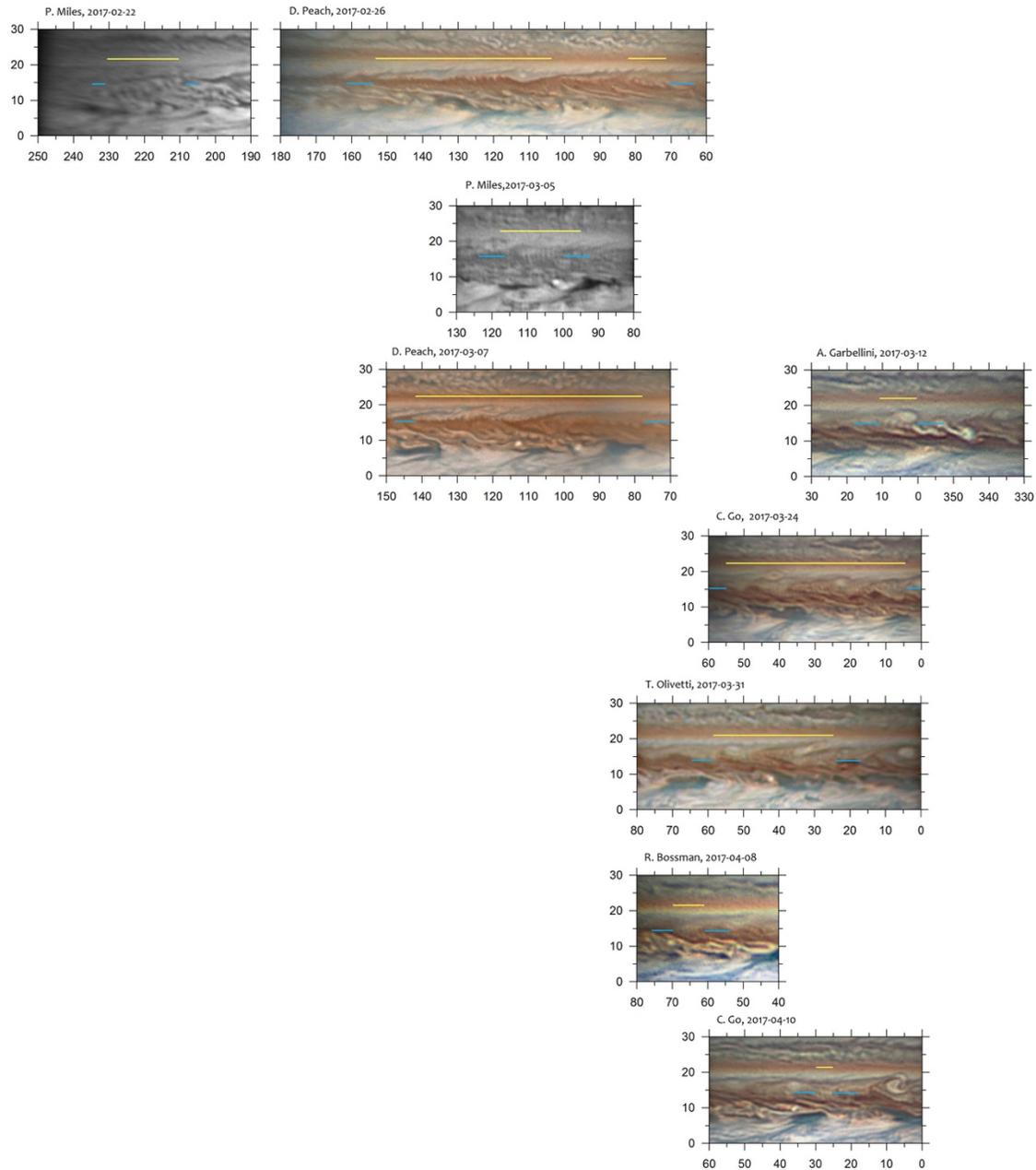

Figure A2. Cylindrical maps of the NEB based on images listed in Table 2. Blue lines show the start and end of the regions where waves are visible. The different wave systems are highlighted with a yellow line above the latitude of interest. All longitudes are given in System III and all latitudes are planetocentric.

Dates are from 22 February to 10 April.

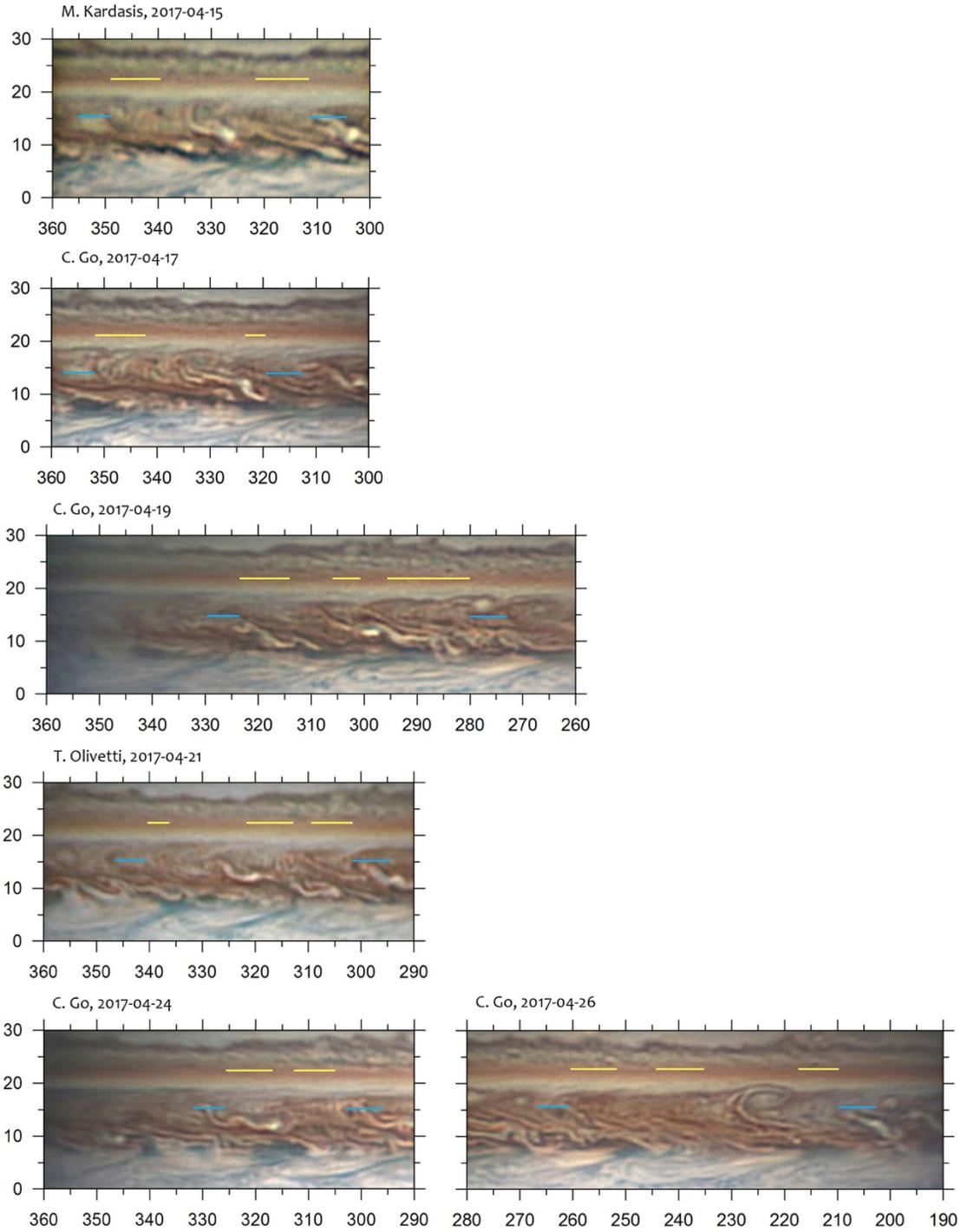

Figure A3: Same as Figure S2 but for dates extending from 16 April to 26 April with more observations due to the proximity to Jupiter opposition on April 6.

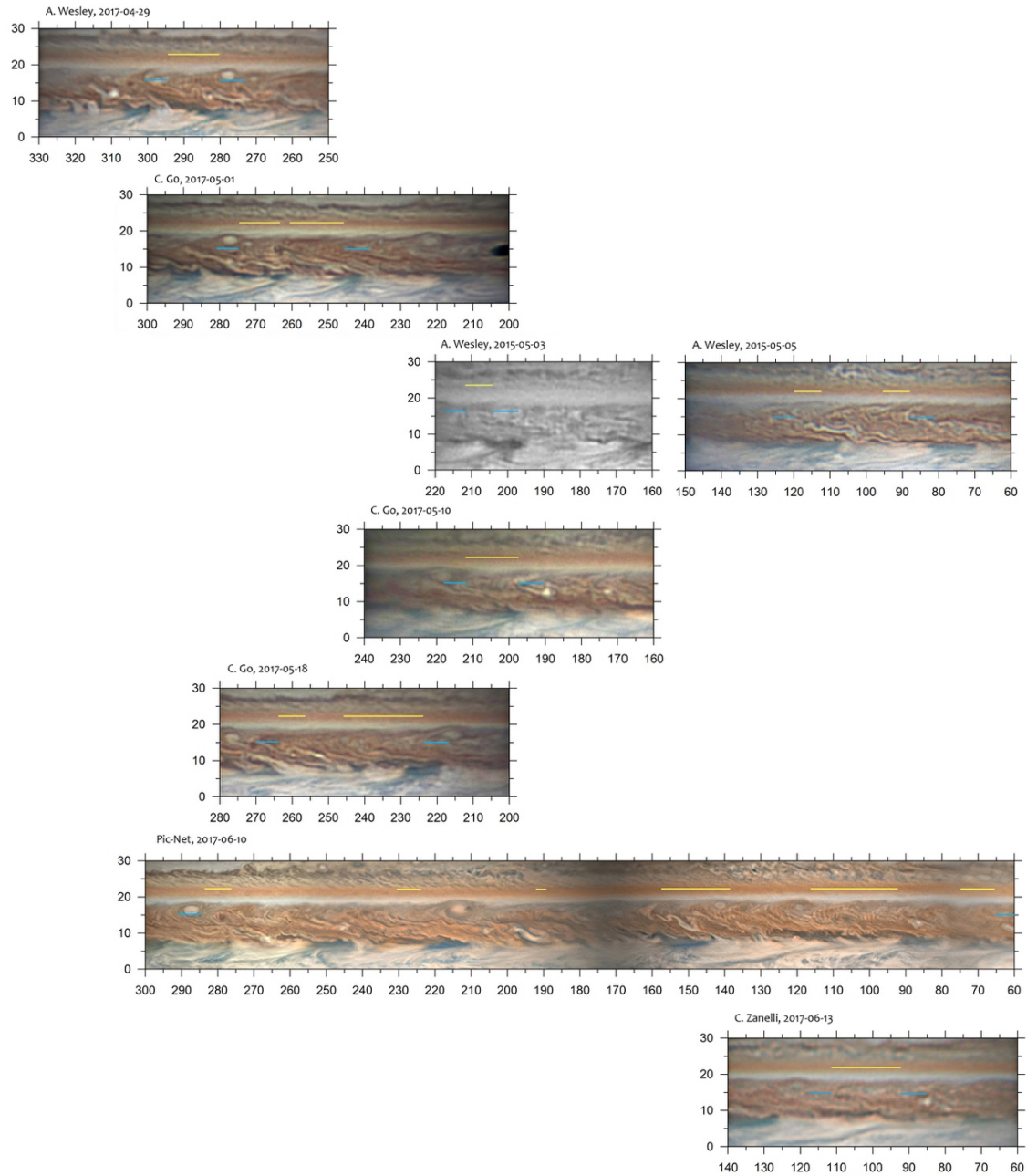

Figure A4: Same as Figure S2 but for dates extending from 26 April to 13 June.

Table A1: Voyager 2 images with identified NEB waves

| Image ID | Start time | Filter |
| --- | --- | --- |
| J_IMG_VG2_ISS_2024224_N | 1979-06-26T00:49:35 | Green |
| J_IMG_VG2_ISS_2024226_N | 1979-06-26T00:51:11 | Violet |
| J_IMG_VG2_ISS_2024228_N | 1979-06-26T00:52:46 | Orange |
| J_IMG_VG2_ISS_2025447_N | 1979-06-26T10:43:59 | Green |
| J_IMG_VG2_ISS_2025449_N | 1979-06-26T10:45:35 | Violet |
| J_IMG_VG2_ISS_2025451_N | 1979-06-26T10:47:10 | Orange |
| J_IMG_VG2_ISS_2026712_N | 1979-06-26T20:39:59 | Green |
| J_IMG_VG2_ISS_2026714_N | 1979-06-26T20:41:35 | Violet |
| J_IMG_VG2_ISS_2026716_N | 1979-06-26T20:43:10 | Orange |
| J_IMG_VG2_ISS_2028950_N | 1979-06-27T14:46:23 | Green |
| J_IMG_VG2_ISS_2028952_N | 1979-06-27T14:47:59 | Violet |
| J_IMG_VG2_ISS_2028954_N | 1979-06-27T14:49:34 | Orange |
| J_IMG_VG2_ISS_2029201_N | 1979-06-27T16:31:11 | Green |
| J_IMG_VG2_ISS_2029203_N | 1979-06-27T16:32:47 | Violet |
| J_IMG_VG2_ISS_2029205_N | 1979-06-27T16:34:22 | Orange |
| J_IMG_VG2_ISS_2032647_N | 1979-06-28T20:19:59 | Violet |
| J_IMG_VG2_ISS_2032649_N | 1979-06-28T20:21:35 | Green |
| J_IMG_VG2_ISS_2032651_N | 1979-06-28T20:23:10 | Orange |
| J_IMG_VG2_ISS_2032917_N | 1979-06-28T22:19:59 | Violet |
| J_IMG_VG2_ISS_2032919_N | 1979-06-28T22:21:35 | Green |
| J_IMG_VG2_ISS_2032921_N | 1979-06-28T22:23:10 | Orange |
| J_IMG_VG2_ISS_2035250_N | 1979-06-29T17:10:23 | Violet |
| J_IMG_VG2_ISS_2035252_N | 1979-06-29T17:11:58 | Orange |
| J_IMG_VG2_ISS_2036413_N | 1979-06-30T02:16:47 | Violet |
| J_IMG_VG2_ISS_2036415_N | 1979-06-30T02:18:22 | Orange |
| J_IMG_VG2_ISS_2036531_N | 1979-06-30T03:19:11 | Violet |
| J_IMG_VG2_ISS_2036533_N | 1979-06-30T03:20:46 | Orange |
| J_IMG_VG2_ISS_2037604_N | 1979-06-30T11:45:35 | Violet |
| J_IMG_VG2_ISS_2037606_N | 1979-06-30T11:47:10 | Orange |
| J_IMG_VG2_ISS_2037616_N | 1979-06-30T11:55:11 | Violet |
| J_IMG_VG2_ISS_2037618_N | 1979-06-30T11:56:46 | Orange |
| J_IMG_VG2_ISS_2037900_N | 1979-06-30T14:06:23 | Violet |
| J_IMG_VG2_ISS_2037902_N | 1979-06-30T14:07:58 | Orange |
| J_IMG_VG2_ISS_2038936_N | 1979-06-30T22:35:11 | Violet |
| J_IMG_VG2_ISS_2038938_N | 1979-06-30T22:36:46 | Orange |
| J_IMG_VG2_ISS_2038940_N | 1979-06-30T22:38:23 | Violet |
| J_IMG_VG2_ISS_2038942_N | 1979-06-30T22:39:58 | Orange |
| J_IMG_VG2_ISS_2038944_N | 1979-06-30T22:41:35 | Violet |
| J_IMG_VG2_ISS_2038946_N | 1979-06-30T22:43:10 | Orange |
| J_IMG_VG2_ISS_2039105_N | 1979-06-30T23:46:23 | Violet |
| J_IMG_VG2_ISS_2039107_N | 1979-06-30T23:47:58 | Orange |
| J_IMG_VG2_ISS_2039109_N | 1979-06-30T23:49:35 | Violet |
| J_IMG_VG2_ISS_2040208_N | 1979-07-01T08:36:47 | Violet |
| J_IMG_VG2_ISS_2040214_N | 1979-07-01T08:41:34 | Orange |
| J_IMG_VG2_ISS_2041654_N | 1979-07-01T20:25:35 | Violet |
| J_IMG_VG2_ISS_2041656_N | 1979-07-01T20:27:10 | Orange |
| J_IMG_VG2_ISS_2042837_N | 1979-07-02T05:47:59 | Violet |
| J_IMG_VG2_ISS_2042841_N | 1979-07-02T05:51:10 | Orange |
| J_IMG_VG2_ISS_2043811_N | 1979-07-02T13:27:11 | Violet |
| J_IMG_VG2_ISS_2043813_N | 1979-07-02T13:28:46 | Orange |
| J_IMG_VG2_ISS_2044103_N | 1979-07-02T15:44:47 | Violet |
| J_IMG_VG2_ISS_2044105_N | 1979-07-02T15:46:22 | Orange |
| J_IMG_VG2_ISS_2044107_N | 1979-07-02T15:47:59 | Violet |
| J_IMG_VG2_ISS_2044109_N | 1979-07-02T15:49:34 | Orange |
| J_IMG_VG2_ISS_2045110_N | 1979-07-02T23:50:23 | Violet |
| J_IMG_VG2_ISS_2045112_N | 1979-07-02T23:51:58 | Orange |
| J_IMG_VG2_ISS_2046324_N | 1979-07-03T09:37:35 | Violet |
| J_IMG_VG2_ISS_2046328_N | 1979-07-03T09:40:46 | Orange |
| J_IMG_VG2_ISS_2047605_N | 1979-07-03T19:46:23 | Violet |
| J_IMG_VG2_ISS_2047607_N | 1979-07-03T19:47:56 | UV |
| J_IMG_VG2_ISS_2047617_N | 1979-07-03T19:55:59 | Violet |
| J_IMG_VG2_ISS_2047619_N | 1979-07-03T19:57:32 | UV |
| J_IMG_VG2_ISS_2047625_N | 1979-07-03T20:02:23 | Violet |
| J_IMG_VG2_ISS_2047627_N | 1979-07-03T20:03:56 | UV |
| J_IMG_VG2_ISS_2047629_N | 1979-07-03T20:05:35 | Violet |
| J_IMG_VG2_ISS_2047631_N | 1979-07-03T20:07:08 | UV |
| J_IMG_VG2_ISS_2047658_N | 1979-07-03T20:28:47 | Violet |
| J_IMG_VG2_ISS_2047700_N | 1979-07-03T20:30:20 | UV |
| J_IMG_VG2_ISS_2047710_N | 1979-07-03T20:38:23 | Violet |
| J_IMG_VG2_ISS_2047712_N | 1979-07-03T20:39:56 | UV |
| J_IMG_VG2_ISS_2047723_N | 1979-07-03T20:48:47 | Violet |
| J_IMG_VG2_ISS_2047725_N | 1979-07-03T20:50:20 | UV |
| J_IMG_VG2_ISS_2047735_N | 1979-07-03T20:58:23 | Violet |
| J_IMG_VG2_ISS_2047737_N | 1979-07-03T20:59:56 | UV |
| J_IMG_VG2_ISS_2047743_N | 1979-07-03T21:04:47 | Violet |
| J_IMG_VG2_ISS_2047745_N | 1979-07-03T21:06:20 | UV |
| J_IMG_VG2_ISS_2047810_N | 1979-07-03T21:26:23 | Violet |
| J_IMG_VG2_ISS_2047812_N | 1979-07-03T21:27:56 | UV |
| J_IMG_VG2_ISS_2047822_N | 1979-07-03T21:35:59 | Violet |
| J_IMG_VG2_ISS_2047824_N | 1979-07-03T21:37:32 | UV |
| J_IMG_VG2_ISS_2047834_N | 1979-07-03T21:45:35 | Violet |
| J_IMG_VG2_ISS_2047836_N | 1979-07-03T21:47:08 | UV |
| J_IMG_VG2_ISS_2048818_N | 1979-07-04T05:32:47 | Violet |
| J_IMG_VG2_ISS_2048822_N | 1979-07-04T05:35:58 | Orange |
| J_IMG_VG2_ISS_2050242_N | 1979-07-04T17:03:58 | Orange |
| J_IMG_VG2_ISS_2050244_N | 1979-07-04T17:05:35 | Violet |
| J_IMG_VG2_ISS_2050258_N | 1979-07-04T17:16:46 | Orange |
| J_IMG_VG2_ISS_2050300_N | 1979-07-04T17:18:23 | Violet |
| J_IMG_VG2_ISS_2050317_N | 1979-07-04T17:31:59 | Violet |
| J_IMG_VG2_ISS_2050331_N | 1979-07-04T17:43:10 | Orange |
| J_IMG_VG2_ISS_2050333_N | 1979-07-04T17:44:47 | Violet |
| J_IMG_VG2_ISS_2051622_N | 1979-07-05T03:59:58 | Orange |
| J_IMG_VG2_ISS_2051626_N | 1979-07-05T04:03:11 | Violet |
| J_IMG_VG2_ISS_2052518_N | 1979-07-05T11:08:47 | Violet |
| J_IMG_VG2_ISS_2052521_N | 1979-07-05T11:11:10 | Orange |
| J_IMG_VG2_ISS_2053924_N | 1979-07-05T22:25:34 | Orange |
| J_IMG_VG2_ISS_2053926_N | 1979-07-05T22:27:11 | Violet |
| J_IMG_VG2_ISS_2053940_N | 1979-07-05T22:38:22 | Orange |
| J_IMG_VG2_ISS_2053942_N | 1979-07-05T22:39:59 | Violet |
| J_IMG_VG2_ISS_2056518_N | 1979-07-06T19:08:46 | Orange |
| J_IMG_VG2_ISS_2056520_N | 1979-07-06T19:10:23 | Violet |
| J_IMG_VG2_ISS_2057557_N | 1979-07-07T03:39:59 | Violet |
| J_IMG_VG2_ISS_2057615_N | 1979-07-07T03:54:23 | Violet |
| J_IMG_VG2_ISS_2058724_N | 1979-07-07T12:49:35 | Green |
| J_IMG_VG2_ISS_2058726_N | 1979-07-07T12:51:10 | Orange |
| J_IMG_VG2_ISS_2058728_N | 1979-07-07T12:52:47 | Violet |
| J_IMG_VG2_ISS_2058730_N | 1979-07-07T12:54:22 | Orange |
| J_IMG_VG2_ISS_2058732_N | 1979-07-07T12:55:59 | Green |
| J_IMG_VG2_ISS_2058734_N | 1979-07-07T12:57:35 | Violet |
| J_IMG_VG2_ISS_2061305_W | 1979-07-08T09:20:45 | CH4_JS |
| J_IMG_VG2_ISS_2061310_W | 1979-07-08T09:24:47 | Orange |
| J_IMG_VG2_ISS_2061315_W | 1979-07-08T09:28:47 | Violet |
| J_IMG_VG2_ISS_2061320_W | 1979-07-08T09:32:47 | Green |